\documentclass[12pt]{article}
\usepackage{epsf,graphics}
\begin{document}
\baselineskip .3in
\begin{titlepage}
\begin{center}
{\large{\bf Temperature Dependent Structure Function of Nucleon }}
\vskip .2in
A. Bhattacharya $^{\ddag}$, A. Sagari, B. Chakrabarti
and S. Mani
\end{center}
\vskip .1in
\begin{center}
Department of Physics, Jadavpur University \\
Calcutta 700032, India.\\
\end{center}

\vskip .3in

\noindent {\bf PACS No.s:} 12.39.-x, 12.39.Jh, 12.39.Pn, 14.20Dh,

\vskip .3in {\centerline{\bf Abstract}}A relation between
temperature, Fermi momentum and radius of the nucleon has been
derived. The percentage increase in radii of the nucleon with
temperature and the temperature dependent F$_{2}$ structure
function of the nucleon have been investigated. An estimate of the
Fermi momentum of the nucleon has been made. The results are found
to be in agreement with existing theoretical and experimental
suggestions . \vskip.1in

$^{\ddag}$ E-mail pampa@phys.jdvu.ac.in
\end{titlepage}

\newpage
{\large{\bf Introduction:-}} \vskip.2in
 The collision of particles provide useful
information about their structure and properties. This method is
widely used in nuclear and high energy physics. There has been
several attempts towards the understanding and estimation of the
structure function of the nucleons [1]. In relativistic heavy-ion
collision, high temperature may be produced by the transforation
of a great deal of energy. So it is important to study the
properties of hadrons at finite temperature. It has been suggested
that the pion cloud contribution plays a crucial role in the long
range structure of the nucleons. Due to its small mass, the pion
plays a special role in the dynamics of hot hadronic matter.
Therefore  it is quite important to understand the temperature
behaviour of the pion's Green function. While the pion is
(hadronically) stable at T=0, it is expected to develop a width
(imaginary part of the Green function) at non zero temperature,
such width being interpreted as a damping coefficient which should
diverge at the critical temperature for de-confinement. This
follows from a proposal [2,3] to consider the width of a hadron as
a phenomenological order parameter for the de-confinement phase
transition. In fact as the temperature is increased and the hadron
melts , its width should increase until it becomes infinite at
T=T$_{C}$ ensuring  no resonance peaks  in the hadronic spectral
function. The latter should become a smooth function of the energy
and coincide with its perturbative QCD value. These properties
have been confirmed by detailed calculation in the framework of
the Linear Sigma Model [4] and the virial expansion [5]. Since,
the discovery of the EMC effect [6], there have been a great
number of discussions on how nuclear structure function vary with
the environment. Most of these discussions relate the structure
functions of a nucleon in a nucleus to the mass number. As the
structure function of a nucleon plays an important role in some
collision process, it is useful to discuss the structure function
of a nucleon at finite temperature. Zhen et al [7] has
investigated the influence of temperature on structure function,
volume energy, radius and mass of the nucleon in the context of
quantum field theory.

In this present work we have derived a relation between
temperature, Fermi momentum and radius of the nucleon in the
framework of the Statistical Model [8]. The increase in radii of
the nucleon with the increase of temperature has been estimated.
The effect of increase in temperature on F$_{2}$ structure
function of the nucleon has been studied. It has been observed
that the Gaussian width of Fermi momentum p$_{F}$=0.54 GeV yields
reasonable value of structure function as predicted by experiment
[6,7].

$\bf{The}$ $\bf{Model}$

 In statistical model [8] the probability
density for a nucleon is obtained as:
 \begin{equation}
 |\Psi(r)|^{2} = \frac{315}{64\pi
 r_{0}^{9/2}}(r_{0}-r)^{3/2}\theta(r_{0}-r)
 \end{equation}
  where $r_{0}$ is the radius parameter of the nucleon and $\theta$ is usual step function.
  The momentum space wave function $\psi(k)$ can be represented as:
 \begin{equation}
 \psi(k) = \frac{c}{k}\int_{0}^{\infty}r\psi(r)dr.sinkr
 \end{equation}
 where c is appropriate normalisation constant. The normalised
 momentum space wave function with (1) is obtained as:
 \begin{equation}
 \psi(k) = 2\sqrt{3\pi r_{0}} k^{-1}j_{1}(kr_{0})
 \end{equation}
 It is to be noted that $\psi(k)$ depends only on the corresponding
 size parameter of the nucleon.

     The free nucleon $F_{2}$ structure function in the
     non-linear limit runs as:
     \begin {equation}
     F(x)=
     \frac{M}{8\pi^{2}}\int_{k_{min}}^{\infty}|\psi(k)|^{2}dk^{2}
     \end{equation}
     where M is the mass of the nucleon and $k_{min}$ = M $|x-
     \frac{1}{3}|$.

 In the statistical model [8] we come across a distribution of ensemble of identical
 indistinguishable Fermions (q and $\overline{q}$) as both the valance and virtual quark are treated at the same footing [8].
 The distribution formula for the Fermions,
 when the temperature is small, (in the first order
 approximation) leads to [9]

\begin{equation}
n_{q}(r)=\frac{N}{V}=\frac{1}{\pi^{2}h^{3}}\int_{0}^{\infty}\frac{p^{2}dp}{1+e^{(\epsilon-\mu)/T}}
\end{equation}

where the chemical potential $\mu$=ar and
$\epsilon$=$\frac{p^{2}}{2m}$, n$_{q}$(r)=number density of quarks
and T is expressed as function of r. The temperature gradient may
be considered as a manifestation of color concentration gradient
from r=0 to r=r$_{0}$ representing the non-equilibrium process in
the context of the model [8] . However we have assumed that the
average state is approximately described by the equations of state
independent of the gradients. To make an estimate of order of the
temperature T, we note that at r=r$_{0}$, n$_{q}$(r$_{0}$)=0 and
at r=0, n$_{q}$(0)= $\frac{315}{64\pi
 r_{0}^{3}}$ so that the average {$\overline{n}_{q}$}=$\frac{315}{128\pi
 r_{0}^{3}}$ and the average chemical potential {$\overline{\mu}$}=$\frac{a r_{0}}{2}$
 as $\mu(0)$=0 and $\mu(r_{0})$=ar$_{0}$. Now integrating the equation (5)
 between p=0 to
p=p$_{F}$, radius of the Fermi sphere in momentum space and
retaining first term we obtain:

\begin{equation}
r_{0}^{4}-\frac{6p_{F}^{2}r^{3}_{0}}{10am}-\frac{945\pi
h^{3}T}{64ap_{F}^{3}}=0
\end{equation}

 With a=0.2 [10], we obtain
 \begin{equation}
r_{0}^{4}-25p_{F}^{2}r_{0}^{3}-\frac{231.82T}{p_{F}^{3}}=0
\end{equation}
Using the value of p$_{F}$ suggested by various models [11] the
structure function of the nucleon has been computed . The value of
$\frac{F_{2}^{T}}{F_{2}^{0}}$ with x at T=0.5T$_{C}$ (with
T$_{C}$= 200 MeV [7,12] and  p$_{F}$=0.54 GeV ) has been computed
and displayed in figure-I. The ratio of the structure function
$\frac{F_{2}^{T}}{F_{2}^{0}}$ of the nucleon corresponding to
different temperature has been estimated and displayed in
figure-II. The dependence of the structure function on p$_{F}$ has
been shown in the figure-III.
 \begin{figure}
\begin{center}
 \rotatebox{-0}{\epsfxsize=7cm\epsfbox{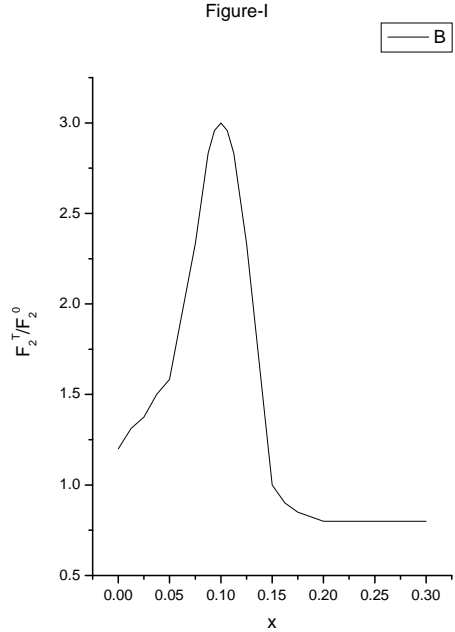}}
\caption{Behaviour of $\frac{F_2^T}{F_2^0}$ with $x$ at $T=0.5T_c$, $p_F=0.54$ GeV}
\label{fig1}
 \end{center}
\end{figure}
\newpage
 {\large{\bf Results and Discussions:-}}
 In the present work we have investigated the influence of
 temperature on the radius of the nucleon and consequently the effect on the structure function has been computed. In the context of the present model
 the structure function has been obtained
  as a function of Fermi momentum. It has been observed that
  as the temperature increases , the radius of the hadron increases until the critical temperature
   but the increase is not very prominent. We have observed that
   at   T=0.5T$_{C}$
 the increase in radius of the nucleon is about 4.56 percent
 and decrease in structure function is 9.46 percent of that at zero temperature
  for x=0.5 whereas Zhen
 et al [7] obtained the increase in radius as 16 percent  and 5 percent decrease in structure function
  at x= 0.3 and T=0.5T$_{C}$. It has been observed that the ratio of the
  $\frac{F^{T}_{2}}{F^{0}_{2}}$ is less than unity at comparatively large x ($\geq$0.2)
  and tends to rise above unity at x ($\sim$ 0.14) at $p_{F}$ =0.54GeV and T = 0.5$T_{c}$.
  It is pertinent to point out here that similar observation has
  been made by EMC [6]
  which confirmed that the nucleon structure function for bound and quasi free
  behaves differently. They have observed that the structure function ratio falls below
  unity at x ($ \geq $ 0.25), rises above unity at x $ \sim$ 0.15 and
  again falls below unity at small x. The variation of the ratio is not very prominent with the variation of the
  temperature whereas Fermi momentum effects the ratio to a
  considerable extent.

  It is interesting to note that Fermi momentum  p$_{F}$=0.54 GeV
 yields the percentage increase in radius and decrease in
 structure function from T = 0 reasonably. It also reproduces the ratio of the structure functions
 in agreement with the EMC [6] predictions. It may be mentioned that $p_{F}$ is a very important parameter
 for extraction of the value of
 CP- violating ratio $|\frac{V_{ub}}{V_{cb}}|$. The precise value  of $p_{F}$ is yet to be determined. Hwang et
 al [11] have extracted the value as the 0.54 GeV in the context of the
 relativistic quark model whereas commonly used value is 0.3 GeV.
 We have observed that $p_{F}$ = 0.54 GeV yields reasonable
 results in the context of the present model. However further
 investigations would be done in our future works.
\vskip .3in
\begin{figure}
 \begin{center}
 \rotatebox{-0}{\epsfxsize=16cm\epsfbox{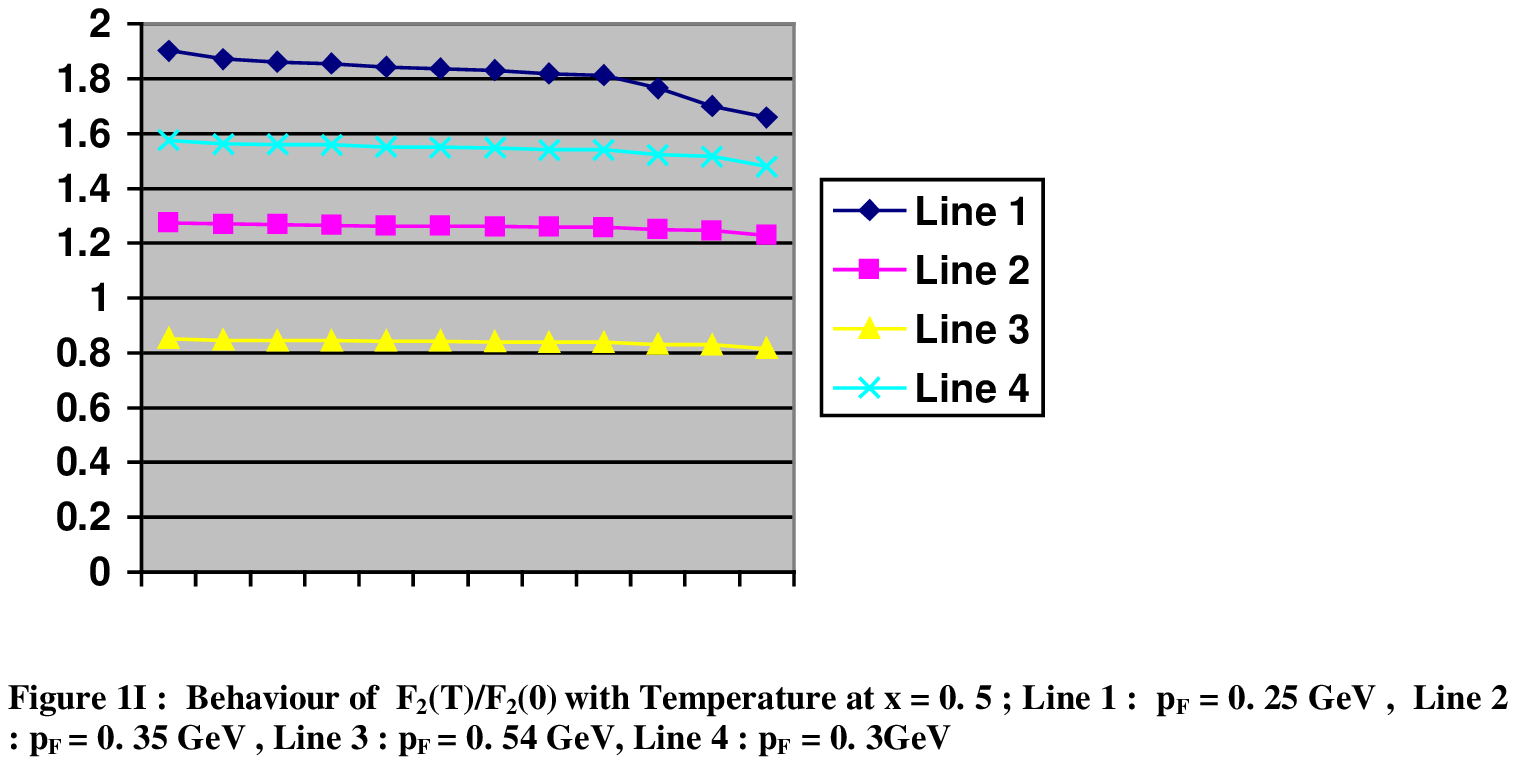}}
\caption{}
\label{fig2}
 \end{center}
\end{figure}
\vskip .3in
\begin{figure}
 \begin{center}
 \rotatebox{-0}{\epsfxsize=7cm\epsfbox{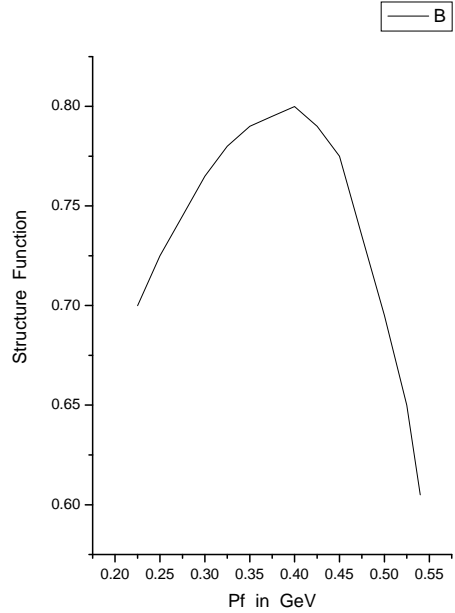}}
\caption{Variation of $F_2$ Structure Function with Fermi momentum $p_F$ in GeV at 
$x=0.5$, $T=0.5T_c$}
\label{fig3}
 \end{center}
\end{figure}
\vskip .3in

\newpage{\bf References}:-

\noindent [1] K.Hagiwara et al., Phys Rev.  D 66(2002) 010001 ;
P.V.Pobylitsa et al., Phys. Rev.  D 59(1999) 034024 ; D.Diakonov
 et al., Nucl.Phys. B 480(1996) 341;

\noindent [2] C.A.Dominguez and M.Loewe, Phys.Lett.B 233(1989)201;
C.A.Dominguez, Nucl.Phys.B(Proc.Suppl.)15(1990)225; C.A.Dominguez
and M.Loewe, Nucl.Phys.B(Proc.Suppl.)16(1990)403; Z.Phys.C,
Particles and Fields,49(1991)423;

\noindent [3] R.D.Pisarski, Phys.Lett.B 110(1982)155;

\noindent [4] C.A.Dominguez, M.Loewe and J.C.Rojas, Z.Phys.C,
Particles and Fields,59(1993)63;

\noindent [5] H.Leutwyler and A.V.Smilga, Nucl.Phys.B
342(1990)302;

\noindent [6] J.Ashman et al.(EMC collaboration) Phys.Lett.B 202
(1988)603;

\noindent [7] B.P.Zhen, J.Phys.G: Nucl.Part.Phys 15(1989)1653;

\noindent [8] A.Bhattacharya et al., Prog.Theor.Phys. {\bf 77}
16(1987), Eur.Phys.J.C. {\bf 2}671(1998), Int.J.Mod.Phys. {\bf A
16}201(2001); S.N.Banerjee et al., Ann.Phys.(N.Y.)150 (1983);

 \noindent [9] B.Chakraborty et al., J.Phys.G 15(1989)L 13 ;

\noindent [10] W.Lucha et al., Phys.Rep. 200,4 (1991) 240 ;

 \noindent [11] D. S. Hwang et al., Z.Phys.C 69(1995)112 and references therein ;

 \noindent [12] D. H. Boal et al., Phys.Rev.D 26 (1982)3285;
\newpage
{\large{\bf Acknowledgement:-}} Authors are  thankful to
Department of Science and Technology (DST), New Delhi,  for
financial  assistance.
\vskip .3in
\noindent
{\bf Figure Captions}:-
\vskip .3in
Figure-I : Behaviour of $\frac{F_{2}^{T}}{F_{2}^{0}}$ with x at
T=0.5T$_{C}$ ,  p$_{F}$=0.54 GeV
\vskip .3in
Figure -II : Calculated values of $\frac{F_{2}^{T}}{F_{2}^{0}}$
with Temperature( T in GeV ) at x=0.5 ; Line 1 : p$_{F}$=0.25 GeV
,Line 2 : p$_{F}$=0.3 GeV , Line 3 : p$_{F}$=0.35 GeV, Line 4 :
p$_{F}$=0.54 GeV
\vskip .3in
Figure -III : Variation of F$_{2}$ Structure Function with Fermi
Momentum ($p_{F}$ in GeV) at x= 0.5, T=0.5T$_{C}$

\end{document}